# A kinetic model of tumor growth and its radiation response with an application to Gamma Knife stereotactic radiosurgery

Yoichi Watanabe[1], Erik L. Dahlman[1], Kevin Z. Leder[2], and Susanta K. Hui[1]

[1]Department of Radiation Oncology
University of Minnesota
420 Delaware St.SE, MMC-494
Minneapolis, MN 55455
612-626-6708
612-626-7060: fax
watan016@umn.edu

[2]Industrial and Systems Engineering
University of Minnesota
111 Church Street SE
Minneapolis, MN 55455
lede0024@umn.edu

Running heads: simple mathematical model of temporal tumor volume change

Contact information

Yoichi Watanabe, Ph.D.
Department of Radiation Oncology
University of Minnesota
420 Delaware St.SE, MMC-494
Minneapolis, MN 55455, USA
Telephone: 612-626-6708
Fax: 612-626-7060
E-mail: watan016@umn.edu

Simple mathematical model of temporal tumor volume change:DRAFT

**Abstract**

We developed a mathematical model to simulate the growth of tumor volume and its response to a single fraction of high dose irradiation. We made several key assumptions of the model. Tumor volume is composed of proliferating (or dividing) cancer cells and non-dividing (or dead) cells. Tumor growth rate (or tumor volume doubling time, $T_d$) is proportional to the ratio of the volumes of tumor vasculature and the tumor. The vascular volume grows slower than the tumor by introducing the vascular growth retardation factor, $\theta$. Upon irradiation the proliferating cells gradually die over a fixed time period after irradiation. Dead cells are cleared away with cell clearance time, $T_{cl}$. The model was applied to simulate pre-treatment growth and post-treatment radiation response of rat rhabdomyosarcoma tumor and metastatic brain tumors of five patients who were treated by Gamma Knife stereotactic radiosurgery (GKSRS). By selecting appropriate model parameters, we showed the temporal variation of the tumors for both the rat experiment and the clinical GKSRS cases could be easily replicated by the simple model. Additionally, the application of our model to the GKSRS cases showed that the $\alpha$-value, which is an indicator of radiation sensitivity in the LQ model, and the retardation factor $\theta$ could be predictors of the post-treatment volume change. Since there is a large statistical uncertainty of this result due to the small sample size, a future clinical study with a larger number of patients is needed to confirm this finding.

Simple mathematical model of temporal tumor volume change:DRAFT

**1. Introduction**

Mathematical modeling of biological processes is widely used to enhance quantitative understanding of bio-medical phenomena. This quantitative knowledge can be applied in both clinical and experimental settings. One important application of modeling exercises is in the area of cancer biology.(Deisboeck *et al.*, 2009; Barillot *et al.*, 2013) Many mathematical models have been developed to represent some aspects of cancer.(Kim *et al.*, 2007; Cristini *et al.*, 2009; Deisboeck and Stamatakos, 2010; Kim *et al.*, 2011) Those models vary from a simple model trying to simulate the growth of tumor volume to sophisticated models including many biologically important molecular processes.(Araujo and McElwain, 2004; Powathil *et al.*, 2013) Macroscopic models can emulate clinically relevant phenomena for anti-cancer therapy. Because of the complexity of the biology, there is a need to link microscopic/molecular processes to those macroscopic ones that we observe in the clinic. Accordingly, there would be a need for a multiscale model rather than just a macroscopic one.

Recently, many investigators began studying mathematical models of tumor response to radiation therapy. The models used for those studies can be categorized into either stochastic or continuum model. Some did extensive studies of the growth and radiation response of the tumor using stochastic models including many biologically important processes at a cellular level, such as apoptosis, angiogenesis, and hypoxia (Borkenstein *et al.*, 2004; Harting *et al.*, 2007; Titz and Jeraj, 2008). Other groups used a continuum model. The tumor growth was represented as diffusion in a three-dimensional space and the model was successfully applied to study the tumor growth and its response to radiation, mainly for brain tumors such as glioblastoma multiforme (GBM) and low-grade glioma (Rockne *et al.*, 2009; Perez-Garcia *et al.*, 2014; Nawrocki and Zubik-Kowal, 2014). There were also several studies in which the non-stochastic modelling approach was chosen to investigate the tumor kinetics after irradiation (Lim *et al.*, 2008; Huang *et al.*, 2010; Chvetsov, 2013; Zhong and Chetty, 2014). In these studies, the tumor volume was represented as a volume in one-dimensional space or the number of cancer cells, and the tumor volume was considered to consist of two components, i.e., proliferating cells and dead cells, after irradiation. Such a model was used to simulate the response of the tumor volume to radiotherapy.

Simple mathematical model of temporal tumor volume change:DRAFT

In this study, we proposed a new simple one-dimensional model for the time evolution of the tumor volume before and after a radiosurgical procedure. We assumed that the tumor volume is composed of actively dividing (or proliferating) cells and non-dividing (or dead) cells after the irradiation. Such a two-component model was originally developed in 1970s (Okumura *et al.*, 1977). This model was previously used to study the radiation response of tumor during radiation therapy (Lim *et al.*, 2008; Chvetsov *et al.*, 2009; Huang *et al.*, 2010). Our model is different from these "standard" models in at least two aspects. First, the slow-down process of tumor growth with increasing volume was modeled by assuming that the growth rate is a function of the ratio of vascular and tumor volumes and the growth rate of the vasculature is slower than that of the tumor itself. Secondly, our radiation response model was created by assuming that the cells do not die instantly, but survive for a few cell cycles. This is supported by some experimental results (Curtis *et al.*, 1973; Forrester *et al.*, 1999; Joiner and van der Kogel, 2009; Sakashita *et al.*, 2013). It is noteworthy that the possibility of one or more mitotic potential of damaged cells was recently considered in a mathematical model of low-grade glioma treated with fractionated radiotherapy (Perez-Garcia *et al.*, 2014) although the authors eventually concluded that this effect is not significant for this treatment. In the current model, the radiation-induced cell kill rate was represented by introducing the probability that the cell stops dividing at the end of the cell cycle. The probability was related to the survival fraction of the linear-quadratic (LQ) model since the probability is not readily available in the current literature.

To test the newly developed model, we exploited the model to reproduce the temporal volume change of rat rhabdomyosarcoma tumors for varying radiation dosages. In 1960s to early 1970s, several British radiobiologists undertook extensive animal studies about the effect of ionizing radiation on tumor volume.(Barendsen and Broerse, 1969; Hermens and Barendsen, 1969; Thompson and Suit, 1969; Tannock and Howes, 1973) These studies provide useful data to test our model. Among those, the article of Barendsen et al (Barendsen and Broerse, 1969) contains experimental data showing the tumor volume before and after a single fraction dose of radiation. Additionally, the model was used to simulate the metastatic tumors of five patients who were treated for their metastatic brain cancers with Gamma Knife stereotactic radiosurgery (GKSRS) technique.

Simple mathematical model of temporal tumor volume change:DRAFT

Our model is simple, but it includes two fundamental processes, the effect of the blood supply to the growth rate and the prolonged mitotic capacity of the cells damaged by radiation, which have not explicitly been included before together within a simple formalism. The applications of the model to clinical cases of GKSRS suggested some of model parameters as a potential predictor of the treatment outcome.

The outline of this paper is as follows. In Section 2, we present the mathematical model with our reasoning behind the selection of the proposed formulae. Numerical solution methods of the set of non-linear ordinary differential equations are briefly discussed in the same section. We used two solution methods. One method employed manual iterative adjustment of the model parameters to fit the model with experimental and clinical data. For another method we applied a simulated annealing technique to automatically estimate the parameters to achieve the best fit between the model and the measurement data. In section 3, we present the sources of tumor volume data, to which our model was fitted. We used the data from old literature of animal experiments and in-house data of five patients treated by GKSRS. The results of the application of the model to those data will be shown in Section 4. After discussion of the models and the results in Section 5, we summarize this study in Section 6.

## 2. Mathematical model

*2.1. Governing equations*

By assuming that the volume of tumor is proportional to the number of cells in the tumor (Ribba *et al.*, 2012), we considered the volume as the basic parameter instead of the tumor mass. We denoted the total volume of malignant cells by $V_T$. A malignant tumor cell divides with a specific cell cycle time, $T_{cc}$. When these cells are exposed to radiation, some of them stop dividing and die. Some fraction of the irradiated cells survives and keeps dividing. The volume of the non-dividing cells is $V_{ND}$. The non-dividing cells are removed from the original site. These processes are represented by the following two ordinary differential equations (ODEs).

$$\frac{dV_T}{dt} = f(V_T, V_v, D)V_T - g(D)V_T \tag{2.1a}$$

Simple mathematical model of temporal tumor volume change:DRAFT

$$\frac{dV_{ND}}{dt} = g(D)V_T - \eta_{cl}V_{ND} \tag{2.1b}$$

The function *f* is the growth rate of the malignant dividing cell, and depends on both the total volume of actively dividing cells, $V_T$, and the tumor vascular volume, $V_v$. The function *g* is the transition rate of the dividing cells to the non-dividing cells under influence of radiation. Both *f* and *g* are a function of dose *D*. The second term in Equation (2.1b) indicates that the non-dividing dead cells are removed away from the original location with the rate of $\eta_{cl}$, which is related to the cell clearance time $T_{cl}$ by $0.693/\eta_{cl}$.

In the current model, the dividing cells are the cells which can divide or proliferate regardless of its clonogenic capacity. Hence, the dividing cells include the cells eventually die due to radiation. Meanwhile, the non-dividing cells are the cells which have stopped dividing and eventually disappear from the tumor volume. Note that we do not consider the cells in a quiescent state such as the cells in the reversible G0/G1 cell cycle arrest state (Schäuble *et al.*, 2012).

*2.2. Growth rate limited by blood volume*

Tumors require new vasculature to provide additional nutrient to grow beyond a certain size, i.e., about 1 mm$^3$. Angiogenesis, the process by which vascular structure develops inside the tumor, is the main mechanism to keep the tumor supplied with sufficient nutrients while it is growing. We represented the amount of the nutrient supplied by the volume of blood or the vascular structure in the tumor, $V_v$. Then, we assumed that the function *f* in Equation (2.1a) is made of two factors and it is given by

$$f(V_T, V_v, D) = \lambda(V_T, V_v)p(D) \tag{2.2}$$

Here, $\lambda$ is the growth rate, which is modulated by the dose-dependent cell proliferation probability *p*. The function $\lambda$ depends on the volumes of both the tumor and the vascular structure. Note that the tumor volume doubling time $T_d$ is defined as $0.693/\lambda$ in this study.

Simple mathematical model of temporal tumor volume change:DRAFT

The growth rate of tumor may depend on the blood volume. Hence, we assumed that the growth rate $\lambda$ is proportional to the ratio of the vascular and tumor volumes:

$$\lambda = a \frac{V_v}{V_T} \qquad (2.3)$$

By taking the time derivative of Equation (2.3), one obtains

$$\frac{d\lambda}{dt} = \left\{ \frac{1}{V_v} \frac{dV_v}{dt} - \frac{1}{V_T} \frac{dV_T}{dt} \right\} \lambda \qquad (2.4)$$

This indicates that the rate of change in $\lambda$ is the difference between the specific volume growth rates of the vasculature and the tumor. For simplicity, we replaced these growth rates with the initial values. Then, we have

$$\frac{d\lambda}{dt} = \{\lambda_v(0) - \lambda(0)\} \lambda \qquad (2.5)$$

We assumed that the volume of the vascular structure increases with the growth rate of $\lambda_v$ as the tumor grows, but the vascular volume grows slower than the tumor:

$$\lambda_v(t) = \theta \lambda(t) \qquad (2.6)$$

The constant $\theta$ is called the retardation factor of the vascular structure (or the vascular growth retardation factor). We considered that $\theta \leq 1$ to realize the saturation phenomena of tumor growth with increasing tumor volume. It is noteworthy that the use of Equations (2.5) and (2.6) along with Equation (2.1a) with only the growth term in the right-hand side of the equation leads to the well-known Gompetzian solution of the tumor growth. See Appendix A.1 for further discussion.

Simple mathematical model of temporal tumor volume change:DRAFT

*2.3. Radiation-induced cell death*

The standard model for radiation-induced cell death after a single dose assumes that a portion of the tumor cells dies and the rest still maintains its proliferation capacity.(Lim *et al.*, 2008; Chvetsov *et al.*, 2009; Rockne *et al.*, 2009; Huang *et al.*, 2010; Rockne *et al.*, 2010) By applying the linear-quadratic (LQ) equation for the cell survival fraction S, then, the volumes of the dividing cells and the non-dividing cells after a single pulse of irradiation at time $t_+$ jump from those at pre-irradiation time $t_-$:

$$V_T(t_+) = S(D) V_T(t_-) \tag{2.7a}$$

$$V_{ND}(t_+) = \{1 - S(D)\} V_T(t_-) \tag{2.7b}$$

where $S(D) = e^{-\chi(D)}$ and $\chi(D) = \alpha D + \beta D^2$. With this model, the function *f* and *g* do not change before and after irradiation and those are given by

$$f = \lambda \tag{2.8}$$

$$g = 0 \tag{2.9}$$

Furthermore, $V_{ND}(t) = 0$ for $t \leq t_-$ for single fraction treatment.

Some studies have shown that cancer cells do not lose their mitotic capability right after exposure to radiation, but those may continue dividing(Curtis *et al.*, 1973; Forrester *et al.*, 1999; Joiner and van der Kogel, 2009; Sakashita *et al.*, 2013). Hence, the volumes of the dividing cells and non-dividing cells gradually change in time. The following radiation response model was formulated by considering this phenomenon. An individual cell divides with probability *p* and does not divide with probability *q* (*=1-p*) in the cell cycle time $T_{cc}$. Then, it is easy to derive the following rate equations for the number of dividing cells, $N_D$, and the number of non-dividing cells, $N_{ND}$:

$$\frac{d N_D}{dt} = \frac{2p - q}{T^*} N_D \tag{2.10a}$$

Simple mathematical model of temporal tumor volume change:DRAFT

$$\frac{dN_{ND}}{dt} = \frac{q}{T^*} N_D \tag{2.10b}$$

Here $T^*$ is a characteristic time, which can be set equal to $T_{cc}$. The initial conditions are

$$N_D(0) = N_0 \tag{2.11a}$$

$$N_{ND}(0) = 0 \tag{2.11b}$$

The solutions of Equations (2.10a) and (2.10b) can be easily found as follows:

$$N_D(t) = N_0 \exp\left(\frac{2p-q}{T^*} t\right) \tag{2.12a}$$

$$N_{ND}(t) = \frac{q}{2p-q} N_0 \left\{ \exp\left(\frac{2p-q}{T^*} t\right) - 1 \right\} \tag{2.12b}$$

We applied these expressions to a radiobiological experiment (Puck and Marcus, 1956), by which one measures the cell survival fraction after a single dose of radiation, *D*. The measured survival fraction *S(D)* is the ratio of the number of dividing (or clonogenic) cells at a specific observation time, or colony counting time, $T_m$, when the cells are irradiated and the number of dividing cells at $T_m$ when the cells are not irradiated.. The data for the surviving fraction is fitted well by the LQ equation(Dale and Jones, 2007; Joiner and van der Kogel, 2009; Hall and Giaccia, 2011). Hence, we obtain the following relationship:

$$\exp\left(\frac{2p-q}{T^*} T_m\right) \exp\left(-\frac{2}{T^*} T_m\right) = S(D) = \exp(-\chi(D)) \tag{2.13}$$

By solving Equation (2.13) for *p*, we obtain

$$p(D) = 1 - \frac{T^*}{3T_m} \chi(D) \tag{2.14}$$

The function *g(D)* is equal to $q/T_m$ :

Simple mathematical model of temporal tumor volume change:DRAFT

$$g(D) = \frac{\chi(D)}{3T_m} \tag{2.15}$$

Note that Equations (2.14) and (2.15) indicate the probabilities $p(0) = 1$ and $q(0) = 0$ when $D = 0$ as expected. For this study, we assumed that the dose is given by a single fraction in duration negligible relative to the time scale in which we monitor the volume variation. Equation (2.14) indicates that the mitotic probability of the cell decreases from unity to a value smaller than unity because of radiation. At the same time, the cells die with the rate of $g(D)$. It was assumed that the effect of the radiation on the cell killing probability lasts for only a certain length, active radiation-effect time, denoted as $\tau_{rad}$. In other words, after the period $\tau_{rad}$, the probabilities $p$ and $q$ are back to 1.0 and 0.0, respectively. It is noteworthy that the total number of proliferating cells dying after irradiation is approximately related to the survival fraction $S(D)$ by

$$\int_{t_-}^{t_- + \tau_{rad}} g(D) N_D(t) dt \approx \{1 - S(D)\} N_D(t_-) \tag{2.16}$$

The number of surviving cells after irradiation of the current model can be the same as the standard model when $\tau_{rad}$ is set approximately equal to $3T_m$. See Appendix A.2 for details.

We assumed that the tumor growth rate is governed by Equation (2.5). During the period of $\tau_{rad}$, we set the right side of this equation to zero; thus assuming that the volume ratio of the vascular structure and the tumor is constant while the cells die due to the radiation effect.

*2.4. Summary of the proposed tumor-volume kinetic model*

In this section we summarized the proposed model used in the rest of this article. Two cell types, or two-components, in tumor, dividing cells and non-dividing cells, were included in the model. Figure 1 illustrates that under the influence of radiation, the dividing cells can continue dividing with the rate of the tumor-growth rate λ(t) times the cell proliferation probability p(D), or can transit to the non-dividing state with a

Simple mathematical model of temporal tumor volume change:DRAFT

rate g(D). The non-dividing cells are eventually cleared from the original location with the cell clearance rate $\eta_{cl}$.

Figure 1: Diagram of the proposed 2-component model

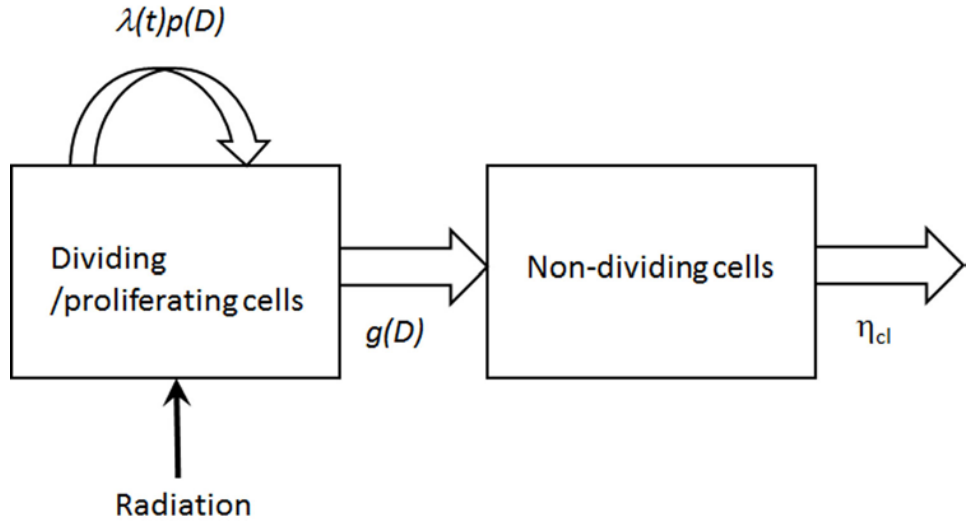

This model can be represented by a set of three equations of three variables: the volume of proliferating tumor, $V_T(t)$, the volume composed of non-dividing cells, $V_{ND}(t)$, and the tumor growth rate $\lambda(t)$, with four constants, a radiobiological parameter $\alpha$, the initial tumor growth rate $\lambda(0)$, the vascular growth retardation factor $\theta$, and the cell clearance rate $\eta_{cl}$.

Initially, the tumor volume grows according to the following ODEs:

$$\frac{dV_T}{dt} = \lambda(t)V_T \tag{2.17a}$$

$$\frac{d\lambda}{dt} = -\theta\lambda(0)\lambda \tag{2.17b}$$

After a single instantaneous irradiation of dose $D$ given to the tumor at $t_R$, the following ODEs are solved for a time period of the active radiation-effect time $\tau_{rad}$, i.e., in $t_R \leq t < t_R + \tau_{rad}$:

$$\frac{dV_T}{dt} = \lambda(t)p(D)V_T - g(D)V_T \tag{2.18a}$$

Simple mathematical model of temporal tumor volume change:DRAFT

$$\frac{dV_{ND}}{dt} = g(D)V_T - \eta_{cl} V_{ND} \qquad (2.18b)$$

$$\frac{d\lambda}{dt} = -\theta\lambda(0)\lambda \qquad (2.18c)$$

For t > $t_R$ + $\tau_{rad}$, $V_T$, $V_{ND}$, and $\lambda$ are the solutions of the following ODEs:

$$\frac{dV_T}{dt} = \lambda(t)V_T \qquad (2.19a)$$

$$\frac{dV_{ND}}{dt} = -\eta_{cl} V_{ND} \qquad (2.19b)$$

$$\frac{d\lambda}{dt} = -\theta\lambda(0)\lambda \qquad (2.19c)$$

Note that three functions, $V_T(t)$, $V_{ND}(t)$, and $\lambda(t)$, are continuous at $t_R$ and $t_R+\tau_{rad}$. The probability p(D) and the transition rate of tumor cells from dividing to non-dividing, g(D), are given as a function of the characteristic time T*, the observation time $T_m$, and $\chi(D)$ by Equations (2.14) and (2.15), respectively. Furthermore, for the LQ model, $\chi(D)$ is a function of $\alpha$ and $\alpha/\beta$ and it is given by(Hall and Giaccia, 2011)

$$\chi(D) = \alpha D \left(1 + \frac{D}{\alpha/\beta}\right) \qquad (2.20)$$

The model parameters $\alpha$, $\lambda(0)$, $\theta$, and $\eta_{cl}$ were obtained to achieve the best fitting of the temporal change of the total tumor volume, i.e., the sum of $V_T$ and $V_{ND}$, between the model and the experimental/clinical data.

For the remainder of this study, we used that $\alpha/\beta$ = 10 days for tumor, the characteristic time T* = 1 day, and the observation time $T_m$ = 10 days. The active radiation effect-time $\tau_{rad}$ is not an optimization parameter but is set to a value between 3 and 10 days. Currently, we do not know well how fast remaining alive cells grow while cells are dying because of irradiation. Hence, it was assumed that $\lambda$ is constant during the time period of $\tau_{rad}$ and it keeps decreasing according to Equation (2.19c) after that.

Simple mathematical model of temporal tumor volume change:DRAFT

*2.5. Solution methods*

A set of nonlinear ODEs, Equations (2.17), (2.18), and (2.19), were numerically solved. The initial conditions were the volume at the first radiology exam for the tumor volume and zero for the volume of non-dividing cells. The initial value of the tumor growth rate was obtained from the tumor doubling time, $T_d(0)$. By introducing a characteristic time that was equal to the cell cycle time and the nominal volume of the dividing tumor cells, those equations were non-dimensionalized for the solution. The solution was obtained by the ODE solver of MATALB, ode15s, which was developed for solving stiff ODEs (The MathWorks Inc., Natick, MA). For the majority of cases in the current study, the parameters were iteratively adjusted so that the temporal change of the tumor volume obtained by the model fits the measured data the best through visual examination.

The parameter estimation was also automatically performed. To fit the mathematical model to the measured tumor volume, we combined a least squares approach with the simulated annealing algorithm (Press *et al.*, 2007). In particular, we needed to compare the N observed tumor volumes denoted by $(y_0,\ldots,y_N)$ and the model predicted volumes at the corresponding times. For a set of parameters $\vec{p} = (\alpha, \theta, T_d, T_{cl})$, we denoted the model predicted volumes at the corresponding times by $(\hat{y}_0(\vec{p}),\ldots,\hat{y}_N(\vec{p}))$. Then our method for finding the best value of $\vec{p}$ was to solve the minimization problem:

$$\min_{p \in \Gamma} \sum_{i=1}^{N} \{y_i - \hat{y}_i(\vec{p})\}^2 \qquad (2.21)$$

Here $\Gamma$ is a set of feasible parameter values.

**3. Experimental and clinical data**

*3.1. Rat rhabdomyosarcoma experiments*

The article of Barendsen et al (Barendsen and Broerse, 1969) contains experimental data showing the tumor volume before and after a single fraction dose of radiation. The dose levels considered were zero (or control), 1000, 2000, 3000, and 4000 rads (or cGy). For their study, rhabdomyosarcoma was grown in rats, which were treated either by 300 kV X-ray or neutron beam. The data are reproduced as Fig. 20.2A in the radiobiology textbook (Hall and Giaccia, 2011). We used the 300kV X-ray data to test our model. It is noteworthy that the tumor volume was measured with a caliper by visually identifying the tumor lesion since no adequate imaging tool was readily available at that time.

*3.2. Clinical examples: metastatic brain tumors treated by GKSRS*

GKSRS is a common modality used to treat intracranial metastatic cancers. For GKSRS, we deliver a large single fraction of the dose to the target by 1.25 MeV gamma rays generated by radioactive cobalt-60 (Co-60) sources. Typical irradiation time is about a half hour, though it depends on the size of the target and the age of the Co-60 source. Typical dosage for malignant metastatic tumors increases as the tumor size decreases (Shaw *et al.*, 2000). GKSRS patients have at least one pre-treatment MR scan. The treatment plan is created with MRI data taken on the day of the treatment. After the treatment, the patients have several follow-up MR scans with the first one, usually, two months post GKSRS. Consequently, during the course of the treatment from the diagnosis to the termination of the follow-up, every GKSRS patient receives several MR scans that use the Gadolinium (Gd) -contrast enhanced T1-weighted imaging technique (CEMRI) with a spin-echo pulse sequence. Therefore, we could collect clinical data showing the change of tumor volumes before and after GKSRS by examining the MRI data. We selected five cases among our patients. These cases were selected so that these can represent typical changes of the tumor volume clinically observed with GKSRS patients. With these clinical data, we investigated if our mathematical model could reproduce the characteristics of the time variation of the observed tumor volume by adjusting the model parameters.

There is a considerable uncertainty about what kind of substance is visualized with CEMRI. It is known that the vasculature of cranial tumors is leaky and the blood-brain barrier (BBB) breaks down (Greenwood, 1991). Hence, the extracellular matrix around the cancer cells can absorb the contrast agent, resulting in a contrast-enhanced volume. Meanwhile, the area around the dead non-dividing cells produced by radiation may or may not enhance on CEMRI. For the current study, we simply assumed that the CEMRI volume is the sum of the dividing and non-dividing cells.

Five clinical cases analyzed for this study are summarized in Table 1. The local control of the tumor after GKSRS are categorized into four groups: complete response (CR), partial response (PR), stable disease (SD), and progressive disease (PD) (Eisenhauer *et al.*, 2009). If the size of the tumor does not change more than 20% after the treatment, it is designated by SD. CR means the disappearance of the tumor. The volume change between CR and SD is considered PR. PD indicates the treatment failure. Using this definition, the local control status of five cases 1 to 5 was assigned to CR, PR, PR, SD, and PD, respectively. The complete response seen in case 1 could be mainly due to the large mean dose (38.3 Gy) given to the tumor in addition to other factors such as the primary tumor type, i.e., non-small cell lung cancer. The failure of the local control of case 5 could be explained by the low maximum dose and the type of the primary tumor. In fact, there is a clinical evidence showing that 20 Gy marginal dose is not sufficient for melanoma metastasis for local control.(Lin *et al.*, 2013).

Table 1: Treatment-related parameters for clinical Gamma Knife stereotactic surgery cases.

| Parameters | Unit | Clinical case number | | | | |
| --- | --- | --- | --- | --- | --- | --- |
| | | 1 | 2 | 3 | 4 | 5 |
| Primary cancer[*] | | NSCL | RCC | RCC | Testicular | Melanoma |
| Prescription dose | Gy | 22 | 16 | 18 | 16 | 20 |
| Prescription isodose line | % | 50 | 50 | 50 | 50 | 80 |
| Maximum dose | Gy | 44.0 | 32.0 | 36.0 | 32.0 | 25.0 |
| Mean dose | Gy | 38.3 | 21.2 | 26.0 | 22.0 | 24.3 |
| Tumor volume at GKSRS | cm^3 | 0.26 | 5.50 | 0.69 | 5.98 | 0.87 |
| Tumor volume at end of follow-up | cm^3 | 0.073 | 0.21 | 0.20 | 4.82 | 23.60 |
| Total monitor duration | days | 137 | 680 | 276 | 106 | 80 |
| Day of GKSRS from initial MRI | day | 34 | 117 | 13 | 78 | 29 |
| Local control[**] | | CR | PR | PR | SD | PD |

[*] NSCL: non-small cell lung cancer, RCC: renal cell carcinoma
[**] CR: complete response, PR: partial response, SD: stable disease, PD: progressive disease.

One of potential applications of a simple model such as ours is to discover a predictor(s) of treatment outcome. Hence, the predictive capability of the proposed model was examined for clinical GKSRS cases in the following way. Using the volume data from the model, first we calculated the change of the tumor volume, $R_{40}$, as the ratio of the volume on the $40^{th}$ day from GKSRS to that at the time of GKSRS. Then, we evaluated the statistical correlation between $R_{40}$ and five variables, i.e., the tumor volume doubling time at GKSRS, $T_d(GKSRS)$, the $\alpha$-value, the $\theta$-value, the tumor volume at GKSRS, $V_{tumor}(GKSRS)$, and the mean target dose, $D_{mean}$. The correlation was analyzed by taking the non-linear regression with the second-order polynomial function between $R_{40}$ and each of the variables. The P-value was estimated for each pair by using the linear regression analysis tool "fitlm", which is available in MATLAB.

## 4. Results

### 4.1. Rat rhabdomyosarcoma experiments

The model parameters of the rat rhabdomyosarcoma experiments are given in Table 2. The initial tumor doubling time, $T_d(0)$, the $\alpha/\beta$ value, and the cell clearance time, $T_{cl}$, were set to the same values for the five cases. The quality of the parameter fitting can be seen in Fig. 2. The experimental data are shown as points in the figure, whereas the volume changes estimated by the model are plotted as solid lines for various dose levels, i.e., 0 (or control), 1000, 2000, 3000, and 4000 rads. The excellent data fitting was achieved by adjusting only the $\alpha$ value and the retardation factor $\theta$ for five different radiation doses. To quantify the magnitude of radiation effect, the survival fraction, S, of the dividing cells at the end of the active radiation period $\tau_{rad}$ was calculated and the values are shown in the lowest row in Table 2. It is noted that the plotting of S as the function of dose generates the typical curve obtained by the LQ model.

The model parameter optimization was also accomplished by using the simulated annealing method for the 3000 rad case. A solution was obtained by selecting the following feasible parameter domain $\Gamma$: $\alpha \in (0.01, 0.2)$, $\theta \in (0, 1)$, $T_d \in (0.5, 15)$, and $T_{cl} \in (1, 50)$. We performed simulated annealing with 100000 steps to numerically solve the minimization problem given by Equation (2.19). The optimal solution was $\alpha = 0.1523$ Gy$^{-1}$, $\theta = 0.7941$, $T_d = 1.5$ days, and $T_{cl} = 5.98$ days. The residual error of this solution was 0.4327; whereas the residual error was 0.5219 when the parameters shown in Table 2 were used. Since the parameter values and the fitting quality are not significantly different and the numerical parameter optimization took much longer computing time, we decided to use the manual iterative method to obtain the optimized values of the model parameters in the rest of the current study.

Table 2: Model parameters for rat rhabdomyosarcoma experiment data: the proposed model$^{(*)}$

| Parameter\Case | Unit | Control | 1000 | 2000 | 3000 | 4000 |
|---|---|---|---|---|---|---|
| $\alpha$ | Gy$^{-1}$ | | 0.30 | 0.20 | 0.16 | 0.145 |
| Dose | Gy | | 10 | 20 | 30 | 40 |
| $T_d(0)$ | days | 1.35 | 1.35 | 1.35 | 1.35 | 1.35 |
| $\theta$ | | 0.72 | 0.72 | 0.74 | 0.795 | 0.838 |
| $T_{cl}$ | days | | 4 | 4 | 4 | 4 |
| MSD$^{(**)}$ | | 0.01883 | 0.01099 | 0.00816 | 0.00832 | 0.03039 |
| S$^{(+)}$ | | 1.00E+00 | 9.83E-01 | 2.84E-01 | 5.38E-02 | 4.51E-03 |

$^{(*)}$ Initial volume = 0.0157cm$^3$. The radiation was turned on at the 11$^{th}$ day. The following model parameters were set constant: $\alpha/\beta=10$ Gy, active radiation-effect time $\tau_{rad} = 8$ days, cell cycle time $T_{cc}= 1$ day, colony counting time $T_m = 10$ days.

(**) Mean square of difference between the experimental data and the model estimated volumes

$^{(+)}$ S is the fraction of surviving cells after irradiation. It is not a parameter used in the simulation model. But, it is included to show the degree of cell kill by radiation. See the text for details.

Simple mathematical model of temporal tumor volume change:DRAFT

Figure 2: Temporal change of the rat rhabdomyosarcoma tumor volume before and after irradiation. The discrete points indicate the experimental data, whereas the solid lines show the tumor volume change produced by the model.

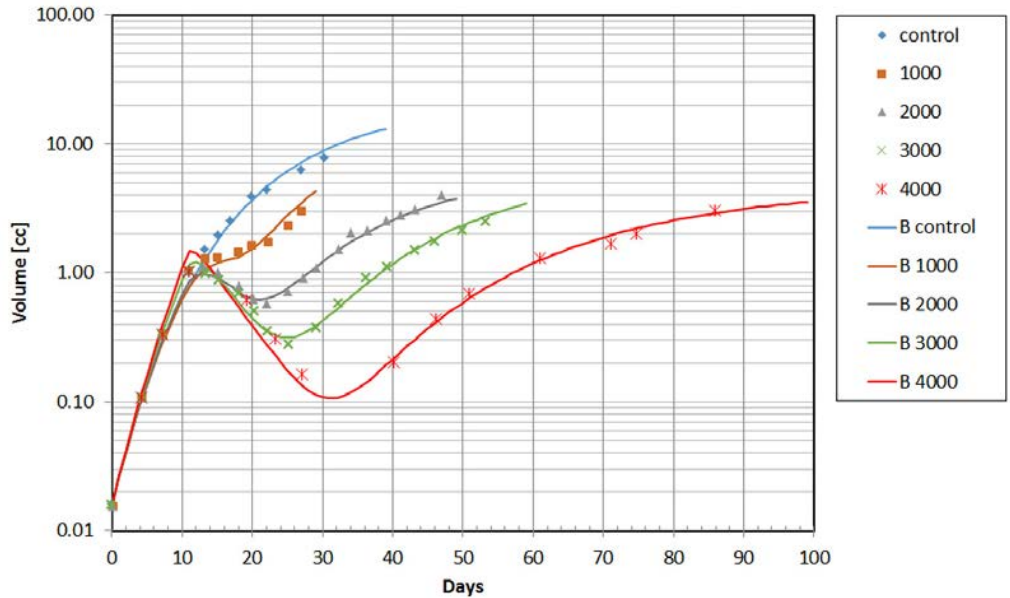

*4.2. Comparison of the proposed model with a traditional model*

In this section we compared the current model (or "proposed model") with more traditional model (or "old model") by applying these models to the data of the rat rhabdomyosarcoma experiment. There are many variations of tumor kinetic models proposed in the literature(Curtis *et al.*, 1973; Okumura *et al.*, 1977; Huang *et al.*, 2010). To simplify the comparison, we used a model in which the Gompetzian growth characteristics is included in the same as the proposed model and the tumor cell death at an the time of instantaneous irradiation is modeled as a sudden change of the volume made of chronogenic cells according to Equations (2.7a) and (2.7b). It is noted that the Gompetzian-type growth characteristics is ignored in many old models, which were used for modeling cell proliferation and radiation effects at the same time.(Okumura *et al.*, 1977)

Simple mathematical model of temporal tumor volume change:DRAFT

For the comparison in this section, we optimized three parameters α, θ, and $T_{cl}$. The final parameter values are given in Table 3. The initial tumor volume doubling time was set to 1.35 days, which is the same as the value used for the proposed model (see Table 2). The temporal changes of the tumor volume obtained by the old model are plotted in Figure 3. It must be emphasized that the non-Gompetzian type model with θ=1 cannot replicate the growth curve of the control case.

There are some distinctive differences between the proposed and old models. The α-value of the old model (0.05 ~ 0.09 Gy$^{-1}$) is much smaller than that of the proposed model (0.145 ~ 0.30 Gy$^{-1}$). The data of the α-value were experimentally obtained for two rat rhabdomyosarcoma cell lines and the value is 0.44 Gy$^{-1}$ for the cells containing H-ras oncogene.(Hermens and Bentvelzen, 1992) Hence, the proposed model predicts the α-value that is closer to the measurement than the old model. Because of the sudden decrease of the tumor volume at the time of irradiation, the old model shows a sharp change of the tumor volume at that time, whereas the prolonged radiation effect used by the current model can smooth out the volume change after the irradiation.

To further evaluate the model fitting capability of the two models, we calculated the mean square of differences (MSD) between the experimental data $\{Y_n\}$ and the volumes estimated by the models at the corresponding time $\{y_n\}$. MSD was defined by calculating a relative error at each measurement point. For $N$ measurement points of data, MSD was given by

$$MSD = \frac{1}{N}\sum_{n=1}^{N}\left(\frac{y_n - Y_n}{Y_n}\right)^2 \qquad (4.1)$$

The MSD values are shown in Table 2 and Table 3. Except the control case, for which MSDs are the same, the MSDs of the proposed model is about a half or less than those of the old model. This clearly indicates a better modelling capability by the proposed model in comparison to the old model.

Simple mathematical model of temporal tumor volume change:DRAFT

Table 3: Model parameters for rat rhabdomyosarcoma experiment data: the old model[(*)]

| Parameter\Case | Unit | Control | 1000 | 2000 | 3000 | 4000 |
|---|---|---|---|---|---|---|
| α | Gy$^{-1}$ | | *0.08* | *0.07* | *0.05* | *0.04* |
| Dose | Gy | | 10 | 20 | 30 | 40 |
| $T_d(0)$ | days | 1.35 | 1.35 | 1.35 | 1.35 | 1.35 |
| θ | | *0.72* | *0.77* | *0.82* | *0.843* | *0.855* |
| $T_{cl}$ | days | | 5 | 5 | 5 | 5 |
| MSD[(**)] | | 0.01883 | 0.02168 | 0.01583 | 0.02722 | 0.05003 |

[(*)] Initial volume = 0.0157cm$^3$. The radiation was turned on at the 11$^{th}$ day. The following model parameters were set constant: α/β=10 Gy, cell cycle time $T_{cc}$= 1 day, colony counting time $T_m$ = 10 days.

(**) Mean square of differences between the experimental data and the model estimated volumes.

Figure 3: Temporal change of the rat rhabdomyosarcoma tumor volume before and after irradiation. The discrete points indicate the experimental data, whereas the solid lines show the tumor volume change produced by the old model.

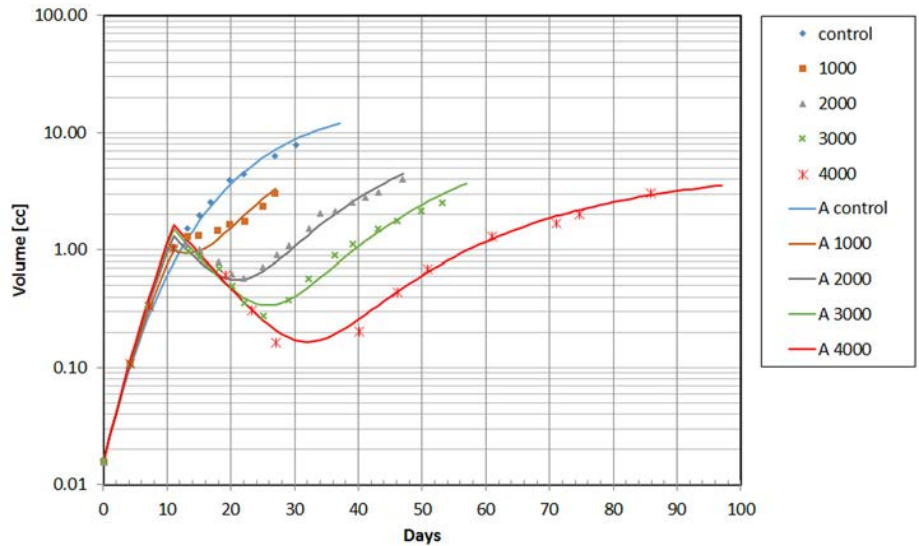

*4.3. Clinical examples: metastatic brain tumors treated by GKSRS*

The clinically observed changes of tumor volumes are presented as open circles in Figs. 4 (a) to (e). The solid lines in the figure indicate the temporal volume change generated by the solutions of the mathematical model developed for this study. Four parameters, $\alpha$, $T_d(0)$, $T_{cl}$, and $\theta$ were optimized for the best fit. Table 4 shows the model parameters used to obtain these curves. Note that the radiobiological parameter $\alpha/\beta$ = 10 Gy, $\tau_{rad}$ = 8 days, and the cell cycle time $T_{cc}$ = 1 day for all cases. We assumed that the number of cell cycles in which the survival fraction was measured to be 10, or $T_m$ = 10 days. The dose used for the simulation was the mean dose prescribed to the target. Due to the technical limitation, the measured volume suffers from a large uncertainty. To assess the uncertainty of the measured tumor volume, hence, we calculated the volume of an equivalent sphere with 0.1 cm expansion or reduction of the radius of the sphere having the same volume as the original volume of the irregularly shaped tumor. Those values are indicated as error bars in Fig. 4.

From the model parameters shown in Tables 1 and 4 and the tumor growth patterns seen in Fig.4, we can make several observations:

a) The range of the $\alpha$ values is between 0.05 and 0.19. These values are on the smaller side of the commonly used range between 0.1 and 0.3 for tumors.

b) The tumor volume doubling time at the time of GKSRS, $T_d(GKSRS)$, ranged from 7.9 days to 32.6 days.

c) The cell clearance time is between 10 and 40 days.

d) The vascular growth retardation factor $\theta$ is larger for the tumors that kept growing even after GKSRS. Particularly, the continued rapid growth after the treatment seen with case 5 required $\theta$ = 0.99, implying that the vasculature grew as rapidly as the tumor volume in this tumor.

The results of the linear regression analyses were shown in Table 5. The P-values of the correlation between $R_{40}$ and five variables, $T_d(GKSRS)$, $\alpha$, $\theta$, $V_{tumor}(GKSRS)$, and $D_{mean}$, were 0.302, 0.026, 0.007, 0.324, and 0.817, respectively. The result indicates significant correlation between $R_{40}$ and $\alpha$ or $\theta$.

Simple mathematical model of temporal tumor volume change:DRAFT

Table 4: Model parameters for radiation response: clinical GKSRS cases[*].

|  |  | Clinical case number | | | | |
| --- | --- | --- | --- | --- | --- | --- |
| Parameters | Unit | 1 | 2 | 3 | 4 | 5 |
| $\alpha$ | 1/Gy | 0.09 | 0.19 | 0.19 | 0.1 | 0.05 |
| Tumor doubling time $T_d(0)$ | days | 29.0 | 6.0 | 9.0 | 6.6 | 7.8 |
| Cell clearance time $T_{cl}$ | days | 13.0 | 38.0 | 40.0 | 10.0 | 20.0 |
| Retardation factor $\theta$ | No dim. | 0.62 | 0.77 | 0.53 | 0.78 | 0.99 |
| Mean dose | Gy | 38.3 | 21.2 | 26.0 | 22.0 | 24.3 |
| Initial volume | cm^3 | 0.126 | 0.0065 | 0.271 | 0.031 | 0.101 |
| Irradiation day for GKSRS | day | 34 | 117 | 22 | 78 | 29 |
| Tumor volume at GKSRS | cm^3 | 0.268 | 0.392 | 0.662 | 6.933 | 1.314 |
| Tumor doubling time at GKSRS | days | 32.6 | 28.0 | 11.2 | 16.1 | 7.9 |
| Volume ratio, $R_{40}$[**] |  | 0.16 | 0.57 | 0.58 | 0.69 | 10.0 |
| % Cell survival fraction |  | 2.24 | 6.03 | 1.99 | 25.4 | 70.0 |

[*] In all cases, $\alpha/\beta = 10$, $\tau_{rad} = 8$ days, $T_{CC} = 1$ day, and $T_m = 10$ days.
[**] $R_{40}$ = the ration of the volume on the 40$^{th}$ day after GKSRS and the volume at GKSRS.

Table 5: P-value for predictors of response $R_{40}$[*]

| Predictor | $T_d$(GKSRS) | $\alpha$ | $\theta$ | $V_{tumor}$(GKSRS) | $D_{mean}$[**] |
| --- | --- | --- | --- | --- | --- |
| P-value: linear | 0.269 | 0.246 | 0.0950 | 0.652 | 0.720 |
| P-value: 2$^{nd}$ order | 0.302 | 0.0264 | 0.0073 | 0.324 | 0.817 |

[*] $R_{40}$ is the ratio of the volume on the 40th day after GKSRS and the volume at the GKSRS, i.e. $V_{tumor}$(GKSRS).

[**] $D_{mean}$ is the mean dose delivered to the tumor for GKSRS.

Simple mathematical model of temporal tumor volume change:DRAFT

Figure 4: Tumor volume as a function of time after the first diagnostic MRI scan. The circles indicate the tumor volume measured on Gd contrast enhanced MRI of GK patients. Solid lines are the volume variation predicted by the model. The vertical arrows indicate the time of GKSRS. The error bars of the measured volumes were obtained by calculating the volume of an equivalent sphere with ±1 mm radius of the original volume. (a) case 1, (b) case 2, (c) case 3, (d) case 4, and (e) case 5.

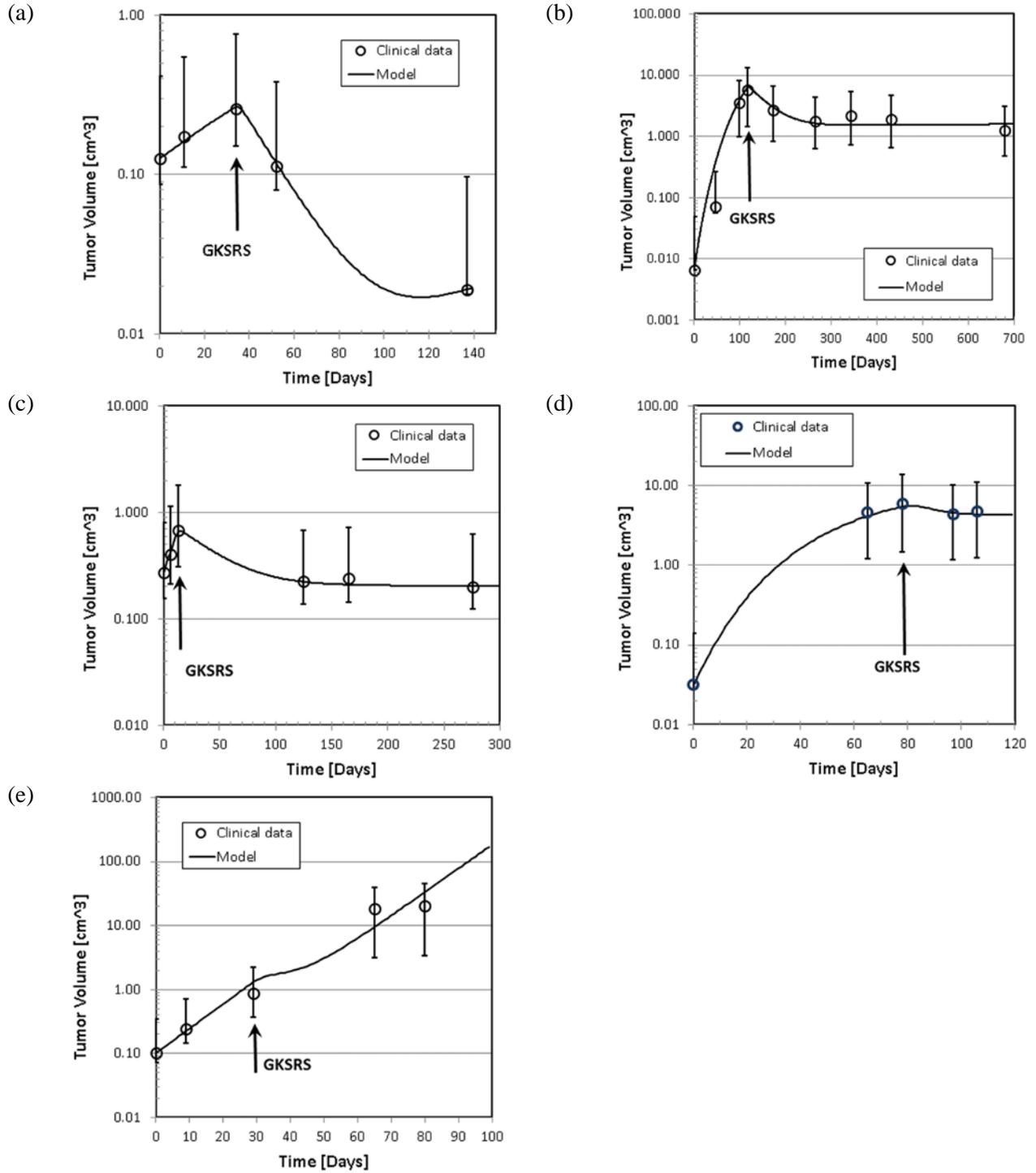

Simple mathematical model of temporal tumor volume change:DRAFT

## 5. Discussion

*5.1. Cell growth model*

It was demonstrated that the proposed model could reproduce the observed growth pattern regardless the initial tumor volumes for the clinical GKSRS cases. The initial volume doubling times of tumors, $T_d(0)$, varied from 6.0 to 29.0 days. The volume doubling time of our model was not constant, but it increased with increasing tumor volume. At the time of GKSRS, the doubling time ranged from 7.9 to 32.6 days. Our own clinical data for the metastatic tumors treated by GKSRS show that the volume doubling time before GKSRS depends on the primary tumor type and the mean values are 42.5, 66.7, and 17.1 days for NSCL, renal cell, and melanoma, respectively (Dalhman and Watanabe, 2012). The doubling time used for our model was within the range of those clinically observed data.

One of the major assumptions made for our model was that the growth rate of tumor volume is a function of the ratio of the volume of the tumor vascular structure to the tumor volume. For simplicity, we assumed that the rate is linearly proportional to the ratio of those volumes. We further assumed that the vascular structure grows with a growth rate proportional to the tumor growth rate. The proportionality constant is named as the vascular growth retardation factor, $\theta$. This factor is considered constant and the $\theta$ value was between 0.53 and 0.99 in the clinical examples. The tumor growth rate, hence, decreases as the tumor size increases. It is noteworthy that the current hypotheses naturally lead to the well-known Gompetzian growth of the tumor volume.

*5.2. Radiation response model*

The mechanism of cell death after irradiation is adequately summarized by an excellent review by B.G.Wouters in Chapter 3 of (Joiner and van der Kogel, 2009). In response to radiation damage, cells lose the replicative capacity and die in genetically controlled mechanisms such as apoptosis (or programmed cell death), autophagy, senescence, and necrosis. These processes take place right at irradiation regardless

Simple mathematical model of temporal tumor volume change:DRAFT

of the cell cycle phase. However, it is considered that a major cell death mechanism in irradiated proliferating cells is mitotic catastrophe, which leads to the cell death due to one of the cell death mechanisms mentioned above, but at the time of mitosis. Hence, this mechanism is also called reproductive or mitotic cell death. Which mechanism is a dominant death process depends on the type of cells, genetic status of the cells, and the microenvironment where the cells reside. Therefore, for the current study, we did not differentiate these mechanisms in the model.

We do not advocate the use of the standard radiation response model for two reasons. First, the survival fraction is a non-dimensional value and the coefficient of the radiation term in the governing equation such as Equation (2.1a) should have the unit of 1/Time. Second, the survival fraction was experimentally obtained by measuring the number of growing colonies in a petri-dish sometime after the irradiation. This time factor is not included in this expression. Hence, we proposed an alternative model, which used a function given by Equation (2.15), instead. The function $g(D)$ is the cell killing probability that depends on the radiation dose and is inversely proportional to the counting time of the surviving colonies, $T_m$, which is the time equal to about ten cell cycles. In addition to the cell killing term, the dividing probability of cell is affected by radiation. This is represented by the probability $p(D)$, or Equation (2.14). After the irradiation, the probabilities $p(D)$ and $q(D)$ remain constant for a certain period. This implies that the cell proliferation capacity is temporarily altered and the tumor cells continuously die during this period.

The probability of a cell to divide after irradiation, p, is represented as a function of dose as given by Equation (2.14). We adopted the LQ formula to relate p with the dose since we do not have alternative data for this relationship. However, for the parameter optimization, p itself can be used. In other words, the LQ model has little meaning in terms of modeling the actual radiation effects. This is one of reasons that the α value of the clinical cases was smaller than the normal range.

Often, tumor volume increases after the treatment. The repair capability of the non-dividing cells, i.e., sublethal-cell repair mechanism (Joiner and van der Kogel, 2009; Hall and Giaccia, 2011), helps to increase the tumor volume after irradiation. However, this repair mechanism is not sufficient to explain the volume growth because the repair takes place within a few hours after the irradiation (Dale and Jones, 2007). The current study showed that the surviving tumor cells after the irradiation can explain the increase in the tumor volume. This phenomenon is often called as "accelerated repopulation". There is another potential mechanism, which helps explain the tumor recurrence. It is known that tumor may be composed of proliferative cells and cells that are slowly dividing. The latter cells termed as "cancer stem cells (CSC)" exist for preserving self-renewal and to protect from external insult such as chemotherapy and targeted therapies.(Dick, 2008; Li and Bhatia, 2011) It is worth pointing out here that there is strong evidence that CSC in glial tumors may grow after irradiation.(Leder *et al.*, 2010; Leder *et al.*, 2014) Explicit modeling of this pathway was not attempted in the current study and the effect of CSC will be investigated in the future.

Our model was successfully fitted to the animal data. The difference among five data sets was only the magnitude of dose that varied from zero to 40 Gy (or 4000 rad). These data contained many temporal data points before and after irradiation. The tumor volume was measured visually with the aid of biopsy; hence, the uncertainty caused by the MRI-based tumor delineation was reduced. Therefore, the application of our model to these data is an excellent method of model testing. The results demonstrated the soundness of the model more clearly than the results seen in the clinical cases.

*5.3. Biological models and parameters*

Our model has similarity to a two-component (or two-compartment) model that was developed in the 1970s(Okumura *et al.*, 1977). The model assumes that the tumor volume is made of cells that are dividing (or proliferating) and cells that do not divide anymore because of fatal radiation damage. The non-dividing cells that can be considered dead cells are transported away from the original location with some time constant, i.e., the clearance time. In the model, the tumor volume is a sum of those two types of cells. Hence, the rate of tumor volume shrinkage is governed by the clearance time. The model has been successfully applied to cervical and head-and-neck cancer cases. (Lim *et al.*, 2008; Huang *et al.*, 2010; Chvetsov, 2013) We also adopted this cell clearance model in this study.

Simple mathematical model of temporal tumor volume change:DRAFT

Recently, the two-component model was extended to a three-component model that includes a reversible state named as "cell-cycle arrest" state and denoted by C (Schäuble *et al.*, 2012) in addition to the proliferating cell (S) and dead cells in an irreversible state (P). Upon responding to external cellular stress, the cells in state P transit to state C. Some fraction of cells in C dies and moves to state S. However, the remaining fraction of cells goes back to P. The transition rates between the states depend on the stress level. The authors looked into the radiation as the stress to validate their hypothetical model. Although the experiments were done in-vitro using human fibroblast cells, their results suggest a possibility of another type of cell repair mechanism in addition to the sublethally damaged cell repair. Our model can be easily extended to test those hypothetical biological mechanisms in the future.

In this study, tumor volume was considered proportional to the number of cells by assuming the tumor cell size does not change during the entire course of treatment. We used the proportionality assumption to estimate the tumor volume change after irradiation by applying the cell survival data of classic radiobiology experiments, i.e., the LQ model. It should be pointed out that there is an experimental evidence which indicates a change in the cell size and the intercellular distance upon irradiation (Tannock and Howes, 1973); thus, leading to a breakdown of our assumption. Considering the strong dependence of such phenomena on the microenvironment and the tumor type, however, we used the current simplifying assumption. The effect of the cell size on the volume change will be studied in the future.

*5.4 Estimated biological parameters*

The rhabdomyosarcoma experiment was done by changing only the radiation dose. However, the best fitting of the model to the experimental data was achieved by varying the $\alpha$-value and the vascular growth retardation factor $\theta$ in addition to the dose. As the dose increased, the $\alpha$-value had to be decreased, whereas the $\theta$-value had to increase as shown in Table 2. Note that the $\alpha$-value is related to the survival fraction of cells, $S$, after single 2 Gy exposure (Dale and Jones, 2007). In fact, $\alpha = 0.145$ and 0.3 with $\alpha/\beta = 10$ lead to S = 71 and 49%, respectively. In other words, a smaller α value implies less effective cell killing capability for the same dose. Thus, our modelling study suggests that the effectiveness of the irradiation decreased as the total dose increased. Note that this result contradicts with the standard LQ model, which would have a difficult time matching the experimental data with a fixed alpha since the dosages are so large. The increase in the θ-value was needed to reproduce the rapid growth of the tumor volume after irradiation. It is true that the θ-value before irradiation should not depend on the radiation dose. However, our model has only one θ-value. In the future model it will be possible to assign different θ-value before and after irradiation for the model consistency.

To accurately model the response to the radiation for the clinical cases, we had to use α values (0.05 to 0.19 $Gy^{-1}$), which are on the smaller side of the commonly accepted range between 0.1 and 0.3 $Gy^{-1}$. The reason for the small α value is not clear, but it may be expected because the mean dose that was used for the modeling study may not be representative for GKSRS due to the highly non-uniform nature of the dose distributions. This warrants further investigation.

*5.5. Delineation of tumor volume*

In this study, we assumed that the volume visible as contrast-enhanced area on T1-weighted MR images (CEMRI) indicates the tumor. There is uncertainty in the volume delineated by this method, however. The Gadolinium contrast permeates into the exogenous and intragenous cellular space by the blood brain barrier disruption caused by the tumors (Roberts *et al.*, 2000). It is a common assumption that the contrasted area contains both the proliferating cells and the dead cells. Hence, we used the sum of the dividing and non-dividing cells as the tumor volume to be compared with the MRI data. It is not clear, however, if this disruption ceases to exist and BBB prevents the contrast material from transporting into the region when the cell dies.

*5.6. Comparison of the proposed model with older models*

The performance of the new two-component model was compared with an older model by applying those to a rat rhabdomyosarcoma experimental data. The new model proposed in this study included a parameter for the slowing-down process of the tumor growth with increasing tumor size (or the Gompertzian-like tumor growth characteristics) and the prolonging effect of radiation after a pulse of irradiation. We found that both models could reproduce the phenomenological characteristics of the growth and radiation response patterns of tumor volume change for varying radiation dosage. However, first it should be reminded that modeling the slowing down of growth is critical to explain the experimental results. Many models currently used in conjunction with radiation response in the field of radiotherapy do not model the Gompetzian-like growth pattern (Lim *et al.*, 2008; Chvetsov *et al.*, 2009; Huang *et al.*, 2010). Secondly, the new model can quantitatively fit the experimental data better than the old model as evidenced by smaller mean square of the differences between modeled volumes and the experimental data. Particularly, the smooth volume change right after a pulse of irradiation can be appropriately modeled only by the new model. Thirdly, one of radiobiological parameters, $\alpha$, estimated by using the old model turned to be too small in comparison to experimentally determined $\alpha$-values for this type of cancer cells. In conclusion, we have demonstrated the superiority of the new model over the old model.

*5.7. Future direction*

Our model contains seven unknown biological parameters, i.e., $\alpha$, $\alpha/\beta$, $T_d$ (tumor doubling time), $T_R$ (cell repair time), $T_{cl}$ (cell clearance time), $\theta$ (vascular growth retardation factor), and the effective time of radiation, $\tau_{rad}$. The radiation dose is known, but there is a possibility to use the dose as an unknown parameter because the physical dose is not uniform inside the tumor volume.

In this study, we showed that there was a significant correlation of the tumor volume change after GKSRS, $R_{40}$, with two model parameters: the $\alpha$-value and the $\theta$-value. Since $\alpha$ indicates the sensitivity of the tumor to radiation, the result is easily predictable. But, the strong correlation between $R_{40}$ and the $\theta$-value is not obvious. This may imply that the tumor responds better to irradiation when the vascular structure grows much slower than the tumor volume. In other words, the shortage of the vasculature slows down or stops the regrowth of the tumor after irradiation. The current results were obtained from only five samples; hence, the confirmation of $\theta$ as a predictor of the GKSRS treatment outcome requires an extensive study with a much larger sample size. Such a future study will enable us to find a correlation between the local tumor control and the model parameters. If these parameters can be diagnostically obtained before the treatment, we should be able to design a patient-specific treatment by adjusting the treatment parameter, that is, the radiation dose to achieve a favorable treatment outcome.

There is experimental evidence showing that the radiation damage to the vascular structure may play a role in enhancing the cell killing capability of radiation for a single large dose of irradiation (Garcia-Barros *et al.*, 2003; Song *et al.*, 2012). For the current study, we assumed the vascular volume, $V_v$, changes along with the tumor volume by assuming its growth rate proportional to the tumor growth rate. The proportional constant was given by the retardation factor, $\theta$. Evidence of this phenomenon was observed with the GKSRS cases we studied as discussed in the above paragraph. To explicitly include the effect within our current modelling framework, we can easily add another equation for $V_v$ and include the radiation damage effect. However, this can be done only when more data on the magnitude of vascular damage as a function of dose will become available.

## 6. Conclusions

We developed a simple mathematical model of tumor growth and its response to radiation by incorporating two key characteristics: (i) the tumor growth rate decreases as the tumor volume increases, and (ii) some radiation-damaged cells still keep dividing for a few more cell cycles after a single pulse of irradiation.

Simple mathematical model of temporal tumor volume change:DRAFT

The current model was successful in mimicking both the animal experimental data and the clinically observed tumor volume changes. We showed that the volume changes of five tumors of Gamma Knife stereotactic radiosurgery patients could be fitted fairly well by using the proposed equations when appropriate model parameters were chosen. A correlation analysis indicated a strong relation between the post-GKSRS tumor volume change and the $\alpha$ and $\theta$-values.

Further refinement of the model which includes the radiation-induced vasculature damage is certainly desirable. Additionally, the model can be easily applied to a larger number of patients treated for GKSRS to find predictors/biomarkers of the treatment outcome. Such a study can confirm the importance of some of the modeling parameters introduced in this study, in particular, the $\theta$-value, to predict better treatment outcome.

**Acknowledgements**

**References**

Araujo R P and McElwain D L 2004 A history of the study of solid tumour growth: the contribution of mathematical modelling *Bull Math Biol* **66** 1039-91

Barendsen G W and Broerse J J 1969 Experimental radiotherapy of a rat rhabdomyosarcoma with 15 MeV neutrons and 300 kV x-rays. I. Effects of single exposures *Eur J Cancer* **5** 373-91

Barillot E, Calzone L, Hupe P, Vert J-P and Zinovyev A 2013 *Computational Systems Biology of Cancer* (Boca Raton, FL: CRC Press)

Borkenstein K, Levegrün S and Peschke P 2004 Modeling and Computer Simulations of Tumor Growth and Tumor Response to Radiotherapy *Radiation Research* **162** 71-83

Chvetsov A V 2013 Tumor response parameters for head and neck cancer derived from tumor-volume variation during radiation therapy *Medical physics* **40** 034101

Chvetsov A V, Dong L, Palta J R and Amdur R J 2009 Tumor-volume simulation during radiotherapy for head-and-neck cancer using a four-level cell population model *Int J Radiat Oncol Biol Phys* **75** 595-602

Cristini V, Li X, Lowengrub J S and Wise S M 2009 Nonlinear simulations of solid tumor growth using a mixture model: invasion and branching. *J Math Biol* **58** 723-63

Curtis S B, Barendsen G W and Hermens A F 1973 Cell kinetic model of tumour growth and regression for a rhabdomyosarcoma in the rat: undisturbed growth and radiation response to large single doses *Eur J Cancer* **9** 81-7

Dale R G and Jones B eds 2007 *Radiobiological Modelling in Radiation Oncology* (Oxfordshire, UK: The British Institute of Radiology)

Dalhman E and Watanabe Y 2012 How fast do metastatic tumors grow in brain? In: *16th International Leksell Gamma Knife Society Meeting,* (Sydney, Austraria: Leksell Gamma Knife Society)

Deisboeck T S and Stamatakos G S eds 2010 *Multiscale Cancer Modeling* (Boca Rayton, FL: CRC Press)

Deisboeck T S, Zhang L, Yoon J and Costa J 2009 In silico cancer modeling: is it ready for prime time? *Nature clinical practice. Oncology* **6** 34-42

Dick J E 2008 Stem cell concepts renew cancer research *Blood* **112** 4793-807

Eisenhauer E A, Therasse P, Bogaerts J, Schwartz L H, Sargent D, Ford R, Dancey J, Arbuck S, Gwyther S, Mooney M, Rubinstein L, Shankar L, Dodd L, Kaplan R, Lacombe D and Verweij J 2009 New response evaluation criteria in solid tumours: revised RECIST guideline (version 1.1) *Eur J Cancer* **45** 228-47

Forrester H B, Vidair C A, Albright N, Ling C C and Dewey W C 1999 Using computerized video time lapse for quantifying cell death of X-irradiated rat embryo cells transfected with c-myc or c-Ha-ras *Cancer Res* **59** 931-9

Garcia-Barros M, Paris F, Cordon-Cardo C, Lyden D, Rafii S, Haimovitz-Friedman A, Fuks Z and Kolesnick R 2003 Tumor Response to Radiotherapy Regulated by Endothelial Cell Apoptosis *Science* **300** 1155-9

Greenwood J 1991 Mechanisms of blood-brain barrier breakdown *Neuroradiology* **33** 95-100

Hall E J and Giaccia A J 2011 *Radiobiology for the Radiologist* (Philadelphia, PA: Lippincott Williams&Wilkins)

Harting C, Peschke P, Borkenstein K and Karger C P 2007 Single-cell-based computer simulation of the oxygen-dependent tumour response to irradiation *Phys Med Biol* **52** 4775-89

Hermens A F and Barendsen G W 1969 Changes of cell proliferation characteristics in a rat rhabdomyosarcoma before and after x-irradiation *Eur J Cancer* **5** 173-89

Hermens A F and Bentvelzen P A 1992 Influence of the H-ras oncogene on radiation responses of a rat rhabdomyosarcoma cell line *Cancer Res* **52** 3073-82

Huang Z, Mayr N A, Yuh W T, Lo S S, Montebello J F, Grecula J C, Lu L, Li K, Zhang H, Gupta N and Wang J Z 2010 Predicting outcomes in cervical cancer: a kinetic model of tumor regression during radiation therapy *Cancer Res* **70** 463-70

Joiner M and van der Kogel A eds 2009 *Basic Clinical Radiobiology* (London, UK: Hodder Arnold)

Kim Y, Magdalena A S and Othmer H G 2007 A hybrid model for tumor spheroid growth in vitro I: Theoreical development and early results *Mathematical Models and Methods in Applied Sciences* **17** 1773-98

Kim Y, Stolarska M A and Othmer H G 2011 The role of the microenvironment in tumor growth and invasion *Progress in biophysics and molecular biology* **106** 353-79

Laird A K 1964 Dynamics of Tumor Growth *Br J Cancer* **13** 490-502

Leder K, Holland E C and Michor F 2010 The therapeutic implications of plasticity of the cancer stem cell phenotype *PLoS one* **5** e14366

Leder K, Pitter K, Laplant Q, Hambardzumyan D, Ross B D, Chan T A, Holland E C and Michor F 2014 Mathematical Modeling of PDGF-Driven Glioblastoma Reveals Optimized Radiation Dosing Schedules *Cell* **156** 603-16

Li L and Bhatia R 2011 Stem cell quiescence *Clinical cancer research : an official journal of the American Association for Cancer Research* **17** 4936-41

Lim K, Chan P, Dinniwell R, Fyles A, Haider M, Cho Y B, Jaffray D, Manchul L, Levin W, Hill R P and Milosevic M 2008 Cervical cancer regression measured using weekly magnetic resonance imaging during fractionated radiotherapy: radiobiologic modeling and correlation with tumor hypoxia *Int J Radiat Oncol Biol Phys* **70** 126-33

Lin H-Y, Watanabe Y, Cho L C, Yuan J, Hunt M A, Sperduto P W, Abosch A, Watts C R and Lee C K 2013 Gamma knife stereotactic radiosurgery for renal cell carcinoma and melanoma brain metastases—comparison of dose response *Journal of Radiosurgery & SBRT* **2** 193-207

Nawrocki S and Zubik-Kowal B 2014 Clinical study and numerical simulation of brain cancer dynamics under radiotherapy *Communications in Nonlinear Science and Numerical Simulation*

Okumura Y, Ueda T, Mori T and Kitabatake T 1977 Kinetic analysis of tumor regression during the course of radiotherapy *Strahlentherapie* **153** 35-9

Perez-Garcia V M, Bogdanska M, Martinez-Gonzalez A, Belmonte-Beitia J, Schucht P and Perez-Romasanta L A 2014 Delay effects in the response of low-grade gliomas to radiotherapy: a mathematical model and its therapeutical implications *Math Med Biol*

Powathil G G, Adamson D J A and Chaplain M A J 2013 Towards Predicting the Response of a Solid Tumour to Chemotherapy and Radiotherapy Treatments: Clinical Insights from a Computational Model *PLoS computational biology* **9** e1003120

Press W H, Teukolsky S A, Vetterling W T and Flannery B P 2007 *Numerical Recipes: The Art of Scientific Computing* (New York, NY: Cambridge University Pres)

Simple mathematical model of temporal tumor volume change:DRAFT

Puck T T and Marcus P I 1956 Action of x-rays on mammalian cells *The Journal of experimental medicine* **103** 653-66

Ribba B, Kaloshi G, Peyre M, Ricard D, Calvez V, Tod M, Cajavec-Bernard B, Idbaih A, Psimaras D, Dainese L, Pallud J, Cartalat-Carel S, Delattre J Y, Honnorat J, Grenier E and Ducray F 2012 A tumor growth inhibition model for low-grade glioma treated with chemotherapy or radiotherapy *Clinical cancer research : an official journal of the American Association for Cancer Research* **18** 5071-80

Roberts T P, Chuang N and Roberts H C 2000 Neuroimaging: do we really need new contrast agents for MRI? *Eur J Radiol* **34** 166-78

Rockne R, Alvord E, Rockhill J and Swanson K 2009 A mathematical model for brain tumor response to radiation therapy *Journal of Mathematical Biology* **58** 561-78

Rockne R, Rockhill J K, Mrugala M, Spence A M, Kalet I, Hendrickson K, Lai A, Cloughesy T, Alvord E C, Jr. and Swanson K R 2010 Predicting the efficacy of radiotherapy in individual glioblastoma patients in vivo: a mathematical modeling approach *Phys Med Biol* **55** 3271-85

Sakashita T, Hamada N, Kawaguchi I, Ouchi N B, Hara T, Kobayashi Y and Saito K 2013 A Framework for Analysis of Abortive Colony Size Distributions Using a Model of Branching Processes in Irradiated Normal Human Fibroblasts *PloS one* **8** e70291

Schäuble S, Klement K, Marthandan S, Münch S, Heiland I, Schuster S, Hemmerich P and Diekmann S 2012 Quantitative Model of Cell Cycle Arrest and Cellular Senescence in Primary Human Fibroblasts *PloS one* **7** e42150

Shaw E, Scott C, Souhami L, Dinapoli R, Kline R, Loeffler J and Farnan N 2000 Single dose radiosurgical treatment of recurrent previously irradiated primary brain tumors and brain metastases: final report of RTOG protocol 90-05 *Int J Radiat Oncol Biol Phys* **47** 291-8

Song C W, Park H, Griffin R J and Levitt S H 2012 *Technical Basis of Radiation Therapy, Medical Radiology, Radiation Oncology,* ed S H Levitt (Berlin: Springer-Verlag) pp 51-61

Tannock I and Howes A 1973 The response of viable tumor cords to a single dose of radiation *Radiat Res* **55** 477-86

Thompson L H and Suit H D 1969 Proliferation kinetics of x-irradiated mouse L cells studied WITH TIME-lapse photography. II *International journal of radiation biology and related studies in physics, chemistry, and medicine* **15** 347-62

Titz B and Jeraj R 2008 An imaging-based tumour growth and treatment response model: investigating the effect of tumour oxygenation on radiation therapy response *Phys Med Biol* **53** 4471-88

Zhong H and Chetty I 2014 A note on modeling of tumor regression for estimation of radiobiological parameters *Medical physics* **41** 081702

Simple mathematical model of temporal tumor volume change:DRAFT

**Appendix A.1: Gompetzian solution**

When there is no radiation, i.e., $D = 0$, Equations (2.1a) and (2.5) in the non-dimensionalized form become

$$\frac{dy_1}{ds} = y_3 y_1 \tag{A.1a}$$

$$\frac{dy_3}{ds} = A y_3 \tag{A.1b}$$

Here, s is the time. $y_1$ and $y_3$ are $V_T$ and $\lambda$ in non-dimension, respectively. By the definition, $A < 0$. It is straightforward to show that the solution for $y_1$ is the Gompetzian function (Laird, 1964):

$$y_1(s) = y_1(0) \exp\left\{ \frac{y_3(0)}{A} [\exp(As) - 1] \right\} \tag{A.2}$$

**Appendix A.2: Estimation of $\tau_{rad}$ from Equation (2.16)**

It is natural to consider that the number of dividing cells, $N_D$, decreases exponentially with the time constant of $g(D)$ after irradiation at time $t$:

$$N_D(t) = N_D(t_-) e^{-g(D)t} \tag{A.3}$$

Plugging Equation (A.3) into the left side of Equation (2.16) and doing the integration, we obtain

$$L.H.S. = N_D(t_-)\left(1 - e^{-g(D)\tau_{rad}}\right) \tag{A.4}$$

Meanwhile, the right hand side of Equation (2.16) is expressed by

$$R.H.S. = N_D(t_-)\left(1 - e^{-\chi(D)}\right) \tag{A.5}$$

By using the formula for $g(D)$ given by Equation (2.15), hence, it is easy to show

$$\tau_{rad} = 3 T_m \tag{A.6}$$

If $T_m$ is set to ten times $T_{cc}$ (= 1 day), $\tau_{rad}$ is 30 days. Note that the value of $\tau_{rad}$ for the clinical cases was set to 8 days as seen in Table 3. This implies that our model resulted in cumulative cell killing less than the LQ model.

Simple mathematical model of temporal tumor volume change:DRAFT

**List of Tables**

Simple mathematical model of temporal tumor volume change:DRAFT**Figure caption**

Figure 1: Diagram of the proposed 2-component model

Figure 2: Temporal change of the rat rhabdomyosarcoma tumor volume before and after irradiation. The discrete points indicate the experimental data, whereas the solid lines show the tumor volume change produced by the proposed model.

Figure 3: Temporal change of the rat rhabdomyosarcoma tumor volume before and after irradiation. The discrete points indicate the experimental data, whereas the solid lines show the tumor volume change produced by the old model.

Figure 4: Tumor volume as a function of time after the first diagnostic MRI scan. The time is in days. The circles indicate the tumor volume measured on Gd contrast enhanced MRI of GK patients. Solid lines are the volume variation predicted by the model. The vertical arrows indicate the time of GKSRS. The vertical arrows indicate the time of GKSRS. The error bars of the measured volumes were obtained by calculating the volume of an equivalent sphere with ±1 mm radius of the original volume. (a) case 1, (b) case 2, (c) case 3, (d) case 4, and (e) case 5.

37